# Hidden photoexcitations probed by multi-pulse photoluminescence

Alexandr Marunchenko[1*], Jitendra Kumar[1], Dmitry Tatarinov[2], Anatoly Pushkarev[2], Yana Vaynzof[3,4], Ivan Scheblykin[1*]

*Corresponding author E-mail: shprotista@gmail.com; ivan.scheblykin@chemphys.lu.se;

[1] Chemical Physics and NanoLund, Lund University, P.O. Box 124, 22100 Lund, Sweden

[2] School of Physics and Engineering, ITMO University, 49 Kronverksky, St. Petersburg 197101, Russian Federation

[3] Chair for Emerging Electronic Technologies, Technical University of Dresden, Nöthnitzer Str. 61, 01187 Dresden, Germany

[4] Leibniz-Institute for Solid State and Materials Research Dresden, Helmholtzstraße 20, 01069 Dresden, Germany

**Abstract.** Time-resolved photoluminescence is a validated method for tracking the photoexcited carrier dynamics in luminescent materials. This technique probes the photoluminescence decays upon a periodic excitation by short laser pulses. Herein, we demonstrate that this approach cannot directly detect hidden photoexcitations with much slower dynamics than the photoluminescence decay. We demonstrate a new method based on a multi-pulse excitation scheme that enables an unambiguous detection and an easily interpreted tracking of these hidden species. The multi-pulse excitation consists of a single pulse (Read) followed by a burst of many closely separated pulses (Write) and finally another single pulse (Read). To illustrate the efficacy of the Read-Write-Read excitation scheme, we apply it to metal halide perovskites to directly visualize carrier trapping, extract the concentration of trapped charges and determine the rate constant of trap depopulation. The developed approach allows studying performance-limiting processes in energy devices using a versatile, highly applicable all-optical method.

# Introduction

Time-correlated single photon counting (TCSPC) is probably the most popular and easily accessible time-resolved optical spectroscopy method[1–3]. It is widely used in fundamental and applied science in areas that span from physics and chemistry to material science, biology and medicine.[1,4–8] The main purpose of TCSPC and other time-resolved photoluminescence (TRPL) methods is to examine the dynamics of various photoexcited species (charge carriers, excitons *etc*) by probing the decay of photoluminescence (PL) generated upon an excitation by a short laser pulse. Quantifying these dynamics is essential for many applications, ranging from photocatalytic energy systems,[9,10] sensors and biological labels[5,8] to optoelectronic devices such as solar cells or LEDs, offering important insights regarding their charge dynamics and energy losses, thus enabling the development of more efficient devices.[6,11,12]

The interpretation of PL decays in samples like solutions of fluorescent dyes can be rather straightforward due to the commonly observed mono-exponential decay kinetics in such systems.[4] However, many other luminescent materials and especially semiconductors, exhibit complex non-exponential PL decay.[1,13–19] The reason for this complexity originates from the interactions between excited states that are able to recombine radiatively (thus giving PL) and other excited species present in the materials, which may impact the concentration of the former. Considering that these excitations (e.g. electrons, holes, excitons, trapped electrons, trapped holes, phonons, defects etc.) participate simultaneously in multiscale and spatially distributed kinetic processes (e.g. recombination, diffusion, trapping, de-tapping, photochemistry),[1,14–18,20–22] the presence of some of these species remains hidden to the traditional PL decay measurements, unless their presence is explicitly considered. In most cases, their influence on the PL is indirect [13,18,23,24] complicating the interpretation of the PL dynamics and quantum yield.

Herein, we discuss the limitations of the conventional time-resolved photoluminescence (TRPL) decay measurements in probing such hidden photoexcited species. We propose a novel time-resolved photoluminescence measurement scheme called Read-Write-Read (RWR) to overcome these limitations. The RWR scheme makes it possible to probe the dynamics at very different time scales and initial conditions within a single experiment. To demonstrate this new method, we apply it to cesium lead tribromide ($CsPbBr_3$) metal halide perovskite microplates as an archetype of a luminescent material with complex photophysics. We show that the RWR method can be used to probe the material state across multiple time-domains ranging from nanoseconds to milliseconds and is an effective tool for the visualization of hidden photoexcited species. In the case of $CsPbBr_3$, the hidden species are trapped charge carriers whose dynamics are quantitatively measured by the RWR experiment. Crucially, the RWR scheme can be readily realized using commercially available equipment designed for traditional TCSPC, making it highly feasible for a board community of physicists, chemists, engineers and material scientists.

## Results and Discussion

**Importance of averaging in TCSPC and the role pulse repetition period**

To understand the principles of operation and the advantages of the RWR technique, it is helpful first to introduce the basics of the conventional time-correlated single photon counting (TCSPC) method, which, while being ubiquitously used, brings forth a number of limitations. Fundamentally, in a TCSPC measurement, the arrival times of PL photons generated by a short excitation laser pulse are measured relative to the arrival time of this excitation pulse (Fig.1a,b).[2,4] Note, that even it is possible, contrary to the common believe, to count several photons per excitation pulse (**Supplementary Note S10**),[1–4] it is still practically impossible to obtain a high signal-to-noise ratio in the PL decay (Fig.1b) when it is excited by just one pulse (**Supplementary Note S1**); thus, it is necessary to repeat the experiment many times to accumulate enough data.

For example, for a laser pulse repetition rate of 1 MHz and a signal acquisition time of 100s, the experiment is repeated 100 million times (Fig.1a). During this acquisition time, the photoluminescence photons are summed up together according to their arrival times relative to the laser pulse they were excited by (Fig.1a,1b). Importantly, this summation does not consider which one of those 100 million pulses excited each particular photon, effectively leading to a loss of this information within the PL intensity histogram, which we illustrate by different colors in Fig.1b. In other words, the resulting averaged PL decay with a high signal-to-noise ratio reflects averaging over the PL dynamics generated by millions of individual laser pulses (Fig.1b).

This repetitive nature of the TCSPC experiment may cause challenges to its interpretation. In particular, one has to be mindful of the fact that in order for the pulse-averaged PL decay to correctly represent the PL decay generated just by one pulse, all of the photoexcited species generated by previous excitation pulses should decay in full prior to the arrival of the next excitation pulse (Fig.1c). Only under these conditions, the PL decay becomes independent on the number of pulses used in the experiment, and thus, can be indeed averaged over their number.

This excitation regime is denoted as a *single pulse excitation regime* (Fig.1c).[20] In the *single pulse excitation regime*, the PL intensity and concentrations of all photoexcited species are periodic functions starting from the very first pulse hitting the sample. Only in this case is the PL decay acquired over millions of periodic laser pulses the same as the one that would have been measured - if it would be technically possible - using a single excitation pulse (Fig.1c).

In contrast, when the decay time of any of the photoexcited species is not negligible in comparison to the pulse repetition period $T$, such long-living excitations created by a single pulse do not completely decay prior to the arrival time of the next pulse (Fig.1d). In other words, the sample holds "memory" of the previous excitation when the next pulse arrives. Consequently, the concentration of these long-living excitations progressively evolves from one pulse to the next (Fig.1d) until it reaches a quasi-equilibrium, denoted as the *quasi-steady-state regime* (Fig.1d).[15,18] Upon reaching this regime, the concentrations of both long-living and short-living photoexcited species again become periodical functions similarly to those in the *single pulse excitation regime* (Fig.1c). However, the concentration of the long-living photoexcited species in this *quasi-steady-state regime* is different to that in the *single pulse excitation regime* (compare Fig.1c and Fig.1d). This quasi-steady-state regime we will call hereafter "*quasi-steady-state regime with memory*" to distinguish from the *single pulse regime*. The term memory highlights that the system remembers the impact of the previous excitations by means of the hidden photoexcited species.

The difference between the two regimes is of great significance for materials whose PL is sensitive to variations in the concentrations of these long-living photoexcited species.[13,23,24,25] Therefore, if one does not consider the pulse-to-pulse dynamics (memory of the previous excitation), only PL decays measured in the *single pulse excitation regime* can be interpreted in a straightforward fashion. The PL lifetime extracted from this PL decay is sometimes referred to as the so-called "true" PL lifetime.

However, even if one knows that TRPL measurements should be done in the single pulse excitation regime to obtain the "true" PL lifetime, how does one know whether or not the specific pulse repetition period falls within the single pulse excitation regime? While it is common to assume that the *single pulse excitation regime* is valid for cases in which the PL has completely decayed (to the noise level) prior to the arrival of the subsequent laser pulse, in the following, we will show that this criterion (from here on referred to as the zero-PL criterion) is not adequate, since in many cases zero PL intensity does not necessarily mean the sample does not contain any photoexcited species (Fig.1d).

**Repetition rate dependent PLQY**

To explore the efficacy of the zero-PL criterion, we first apply the standard TRPL approach with a variable pulse repetition rate from 4 kHz to 2 MHz (see also **Supplementary Note S1**) on a reference dye solution (Fluorescein 27) and on $CsPbBr_3$ perovskite microplates, whose synthesis was reported earlier[26] (see also **Supplementary Note S2**). As shown in Fig. 2a,b, for both types of samples, the PL intensity decays to the noise level within several tens of nanoseconds. This means that within the entire range of pulse repetition periods (from 500 ns to 250 μs), the commonly used zero-PL criteria are satisfied by a large margin.

However, these two samples still show drastically different responses to a change in the repetition period. While the PL signal for the reference dye remains nearly identical (less than 10% change in the initial amplitude) (Fig. 2a), the PL signal measured from the $CsPbBr_3$ microplates markedly increases with the increasing pulse repetition rate. Considering that the samples were always excited by the same number of pulses (assured by keeping the acquisition time of each experiment inversely proportional to the pulse repetition rate), the measured PL is directly proportional to the PLQY. Consequently, while the reference dye preserves a fixed PLQY, the PLQY of the $CsPbBr_3$ microplates is enhanced by a factor of 18 upon increasing the repetition rate from 4 kHz to 2 MHz (Fig. 2b,c).

From Fig. 2c we infer that for the pulse repetition rates equal to and below 20 kHz (equivalent to 50 μs repetition period between pulses, Fig.2c), the sample is likely in a *single pulse excitation regime* because the PLQY in this repetition range remains unchanged.[20] However, as the repetition rate exceeds 20 kHz, the PLQY starts to be dependent on this repetition rate. It means that the distance between the pulses is not long enough to ensure the decay of all photoexcited species. Note that this effect is not reflected at all in the PL decays while the zero-PL criteria is always valid (Fig.2c).

Taken together, these results demonstrate that the standard TRPL measurement fails to directly probe the full picture of the photoexcited carrier dynamics in the presence of hidden photoexcitations. The standard TRPL approach can primarily suggest the presence of hidden photoexcitations by changes in absolute PLQY and lifetime at different repetition rates (Fig. 2c,d). However, this method shows only the PL governed by the equilibrated and averaged charge carrier dynamics, under quasi-steady-state conditions (Fig. 1c,d). To detect hidden photoexcitations, one would need to apply a theoretical model and numerically solve these dynamics from the experiment's start to end.[18,20,24,25] Therefore, this approach will always be constrained by the initial theoretical assumptions. Moreover, the standard TRPL approach might prove challenging to apply to highly sensitive luminescent materials such as metal halide perovskites as it would require multiple measurements at different repetition rates. The changes in the sample properties of sensitive materials (e.g. due to degradation) might impact these measurements, thus introducing additional processes that would not be reflected in the theoretical model.

**TCSPC with Pulse Burst Excitation**

dynamics can be directly visualized if one applies non-standard excitation and detection schemes.[23,24,27,28] Specifically, rather than applying a standard periodic pulsed excitation, we excite the $CsPbBr_3$ sample with pulse bursts repeated with a very long repetition period (for more details, see **Supplementary Note S1**). We realized this excitation scheme using rather standard TCSPC electronics, as illustrated in Fig. 3a,b.

Fig.3c shows the response of the $CsPbBr_3$ sample to a pulse burst consisting of 1000 pulses (Fig.3d) with an internal repetition rate of 80 MHz (i.e. 12.5 ns between pulses) and a period of T = 1 ms between the pulse bursts. This long burst repetition period (repetition rate 1 kHz) is chosen to enable the sample to return to its true ground state (the ground state for all relevant excited species) prior to the arrival of the next excitation pulse burst. The pulse burst experiment makes it possible to simultaneously track the evolution of both the relative initial PL amplitude and PL lifetime after each pulse (inset in Fig. 3c).

Let us interpret this multi-pulse response in terms of the *single pulse* and *quasi-steady state excitation* regime with the memory discussed above. The very first pulse of the burst excites the sample in its true ground state. This means that the PL response to the first pulse is excited in the *single pulse excitation regime*. Thus, it should be identical to the PL decay generated by usual TCSPC at 20 kHz repetition rate and lower (Fig. 3c).

Later, each new pulse increases the concentration of the hidden photoexcited species because of the presence of memory between the pulses, generating more and more intensive PL until it reaches a quasi-steady-state regime close to the end of the pulse burst. Therefore, the last pulses of the pulse burst give a similar PL response as measured by the TCSPC at the 80 MHz repetition rate.

Thus, the whole measured PL response shows the conversion from the *single pulse excitation regime* to the *quasi-steady state excitation regime with memory*. Moreover, it also shows the transition between the two, which contains information about the dynamics of the hidden excitations.

**The Read-Write-Read TRPL technique.**

To further enhance the capabilities of the pulse burst measurement scheme, we propose a scheme called Read-Write-Read (RWR) TRPL (Fig.4). In the RWR TRPL the sample is exposed to the following repeated pulse pattern: a single pulse (denoted $Read_1$), a delay time $t_1$, a burst of N pulses (denoted Write, with a total length of $t_2$ and a delay time between pulses of $T_W=t_2/(N-1)=1/f_W$, where $f_W$ is the pulse repetition rate within the pulse burst), a delay time $t_3$, and finally one more pulse (called $Read_2$) followed by a delay time $t_4$. The total repetition period of the RWR excitation sequence is $T=t_1+t_2+t_3+t_4$.

To illustrate the RWR TRPL we modelled this experiment for a semiconductor where the hidden photoexcited species are trapped charge carriers. Details of the modelling can be found elsewhere[20,24] and in **Supplementary Note S3**. Accumulation of trapped carriers creates the so-called photodoping effect that indirectly but significantly influences the PL. The top panel of Fig.4 shows the calculated PL response to the RWR pulse excitation sequence shown in bottom panel. To understand this response, it is important to compare the PL intensity (red curve) with the concentration of the trapped carrier population (hidden photoexcited species, violet curve). This comparison also makes it possible to understand the meaning of each of the experimental parameters within the RWR experiment, as discussed below:

- The pulse sequence begins with the $Read_1$ pulse, which creates the initial PL decay (Fig. 4). In case of a sufficiently long repetition period T, the PL decay is collected in the *single pulse excitation regime*. Hence, this PL decay is similar to the one obtained in the conventional TRPL experiment with a very low pulse repetition rate (very long pulse repetition period). Therefore, this PL decay can be used as a reference for comparison with the other PL decays generated later by the RWR excitation pattern.

- Then, the delay time $t_1$ is applied to prevent the PL decay initiated by $Read_1$ from overlapping with that generated by the Write burst that would follow.

- Next, the Write pulse burst is applied in order to substantially change the state of the sample in terms of the concentration of photoexcited species. The repetition rate inside the Write burst ($f_W$) should be high enough to induce clear pulse-to-pulse PL dynamics (due to hidden photoexcited species) (Fig. 1d, Fig.3c, Fig.4). With each additional pulse within the Write burst, the population of the photoexcited species in the sample is changed (Fig. 4). A large enough number of applied pulses ultimately shifts of the excited system to a *quasi-steady-state regime* (Fig. 4). Thus, each individual PL response at the end of such a long Write burst should be the same as the one measured using the conventional TRPL experiment at the repetition rate $f_W$.

- Next, after a $t_3$ delay time, a Read$_2$ pulse is applied in order to probe the dynamics of the hidden photoexcited species. A comparison between the PL responses to the Read$_2$ pulse versus that to the Read$_1$ pulse provides information about the hidden photoexcited species impacting the PL and still present at the time $t_3$. Therefore, by varying the delay time $t_3$, it is possible to resolve the evolution of the hidden photoexcited species in time.

- Finally, the delay time $t_4$ has to be very large, generally $t_4 >> t_1, t_2, t_3$. This condition is necessary for the sample to relax back to the ground state before the next RWR pulse sequence commences. This is because - similarly to the single pulse TRPL - the RWR TRPL also requires averaging over many RWR cycles to achieve a high signal-to-noise ratio.

Therefore, in contrast to the conventional TRPL technique, in which the repetition period T is the only relevant timescale, the RWR TRPL technique includes multiple timescales (Fig. 4) controlled by the choice of the five parameters: $t_1$, N, $f_W$, $t_3$, and $t_4$. The multiparameter nature of the RWR methodology is key to its ability to measure both slow and fast dynamics of the photoexcited carriers.

**Visualizing trap population and depopulation in CsPbBr$_3$ microplates**

Studying the PL decay in metal halide perovskites (MHPs) is important not only for a fundamental understanding of their electronic properties but also for the characterization and optimization of photovoltaic and optoelectronic devices based on these semiconductors.[6,12] Recent reports showed that the timescale of the slow charge carrier dynamics in MHPs may lie in the range of tens and even hundreds of $\mu$s and is related to the relaxation of charge carriers trapped in defect states.[23,24,29–31] At the same time, because the PL decays in MHPs rarely exceed 1 $\mu$s, the trapped charge carriers in MHPs can be considered as hidden photoexcited species.

To probe the dynamics of trapped charges in CsPbBr$_3$ microplates, we applied the RWR TRPL technique with the parameters of the excitation shown in Fig.5a to obtain the PL response shown in Fig.5b. Note that the PL lifetime of these CsPbBr$_3$ microplates is around 1-2 ns only (Fig. 2b). This is much shorter than the time between two subsequent pulses (12.5 ns, $f_W$=80 MHz) within the Write burst. This makes the PL signal very low before each subsequent pulse within the Write burst (see also **Supplementary Note S4**). Nevertheless, the PL generated by each incoming pulse within the Write burst steadily increases. This makes the integrated PL response grow by approximately 30 times at the end of the Write burst relative to the PL response obtained from the Read$_1$ pulse.

This enhancement occurs due to the so-called photodoping effect caused by the accumulation of long-living trapped charge carriers. [1,20,23,24,32–34] If, for example, electrons quickly reside in traps, then the concentration of free holes (p) becomes larger than the concentration of free electrons (n) by the concentration of trapped electrons ($n_t$), because of the charge neutrality condition (n + $n_t$ = p). Therefore, a large PL enhancement within the Write burst means a large photodoping.

We additionally can probe the difference between the PL response to Read$_2$ pulse at delay time $t_3$ and the PL response to Read$_1$ pulse, which shows the impact of the trapped charges (hidden photoexcited species) existing in the perovskite at time t₃ on PL. Thus, to get information about the trap state population decay, we tuned the delay time t₃ while measuring the PL response to the Read₂ pulse (see **Supplementary Note 4**). The ratio of the integrated response PL(Read₂) to that of the reference PL(Read₁) is presented in **Supplementary Note 5**. This dependence shows the decay of the memory in the sample created by the Write burst. It is rather complex and can be fitted by a three-exponential function with the characteristic timescales of 0.38 µs, 2.1 µs and 25 µs. As one would expect, the longest timescale matches the estimate of the pulse repetition period in the classical TRPL method, which ensures the sample is in the *single pulse excitation regime*.

Because in metal halide perovskites, the hidden photoexcited species are trapped charge carriers (21,22), we can use the Shockley-Read-Hall (SRH) recombination model (**Supplementary Note S3**) (18,22) to explain the RWR experiment on CsPbBr₃ microplates quantitatively. We can obtain some quantitative results analytically for the simplest SRH model with one trap type. The model contains three types of photoexcited species: electrons at a concentration n(t), holes at a concentration p(t) and filled traps (or trapped electrons) at a concentration $n_t$(t) [20,23,24]. The PL intensity in such a system is given by:

$$PL(t) = A k_r n(t) p(t) \quad (1)$$

where A is a constant determined by the setup's light collection and detection efficiencies, and $k_r$ is the radiative recombination constant.

Prior to the arrival of the Read₁ pulse, the sample is in the ground state, meaning that all the above-listed concentrations equal to zero (Fig. 5d). The first excitation pulse (Read₁ in our scheme) at the moment of its absorption creates $n_0$ photoexcited electrons and holes (n=p=$n_0$). Therefore, the initial PL amplitude (PL$_{in}$) equals to:

$$PL_{in}(Read_1) = A k_r n_0^2 \quad (2)$$

This equation gives the initial PL amplitude in the *single pulse excitation regime*, and we use this response as a reference.

Applying the charge neutrality condition (p=$n_t$+n) and using Eq. 1 and 2 we can express the initial PL amplitude created by any excitation pulse exciting the system which is not in the ground state in terms of charge carrier concentrations:

$$PL_{in}(pulse) = A k_r (n_0 + n)(n_0 + n_t) = PL_{in}(Read_1) + A k_r (n_0 n_t + n n_0 + n\, n_t) \quad (3)$$

Where n and $n_t$ are the concentrations of free and trapped electrons just before the arrival of the excitation pulse. These concentrations reflect the memory held by the semiconductor from the previous excitations.

At the condition of low PL quantum yield and strong photodoping (n<<p, $n_t$), the PL decay Eq.1) reflects the decay of the minority charge carriers (electrons in our case, see **Supplementary Note 6**). Thus, if the PL intensity generated by the pulse of interest is much larger than the residual PL present just before the arrival of this pulse, then n <<$n_0$. All these simplifies Eq.3 to:

$$PL_{in}(pulse) = PL_{in}(Read_1) + Ak_r n_0 n_t \qquad (4)$$

and in combination with Eq.2 we obtain:

$$n_t = n_0 \left( \frac{PL_{in}(Pulse)}{PL_{in}(Read_1)} - 1 \right) \qquad (5)$$

Because in the experiment shown in Fig.5b the PL intensity substantially decays before each new pulse of the burst arrives, Eq.5 is valid within the Write burst. This means that the growth of the initial amplitude of PL above the initial amplitude of the reference response ($PL_{in}(Read_1)$) shown in Fig.5b is directly proportional to $n_t$.

All of the discussed above are also applied when we probe the system by the $Read_2$ pulse. So, the decay of $n_t$ created by the Write burst can be directly obtained from the initial amplitude $PL_{in}(Read_2)$ using Eq.5. (Fig. 5c). This graph shows that the concentration of the trapped carriers by the end of Write is 16 times larger than $n_0$=1.33×10$^{15}$ cm$^{-3}$ and equals to $n_t^0$ = 2.2×10$^{16}$ cm$^{-3}$. From this value, we also conclude that the concentration of defect states (trap density) responsible for photodoping must be at least larger than 2.2×10$^{16}$ cm$^{-3}$ in the CsPbBr$_3$ microplates used here.

Moreover, in the framework of the SRH model used here, and at the condition of strong photodoping (n<<p≈$n_t$), the decay of $n_t$ can be obtained analytically (see **Supplementary Note 7**):

$$n_t = \frac{n_t^0}{(1 + k_n n_t^0 \, t)} \qquad (6)$$

Fig.5c shows that the experimentally obtained dependence of $n_t$ can be very well approximated by Eq.6, allowing us, for the first time, to directly extract the trap depopulation rate constant $k_n$ = 1.5×10$^{-10}$ cm$^3$s$^{-1}$ from TRPL measurements. Quantifying the trap depopulation constant is important to the design of new strategies to control defects in electronic and optoelectronic applications such as perovskite field-effect transistors, light-emitting diodes, photodetectors and solar cells since traps and their dynamics strongly impact device performance in such devices.[35–38]

The presence of long-living hidden excited species enables the sample to "remember" the history of its previous excitation.[23,24,39,40] Indeed, as discussed above, the PL intensity generated by a $Read_2$ pulse depends not only on the parameters of the pulse itself but also on how often and for how long the sample was excited before. One can see this as a memory effect which can be read by the PL. This concept is the foundation of a novel device called memlumor (luminophore with memory), whose excitation history-dependent PL quantum yield is not a constant but a function of internal parameters determining the state of the sample [23]. Thus, we believe that the RWR TRPL method will become an essential instrument in the emerging field of luminescent devices such as memlumors. [23,24,39–42]

While other, more complex techniques exist for probing the populations of trapped charges (see **Supplementary Note 8** for details), the RWR TRPL offer a clear advantage due to its simple and inexpensive experimental realization. Commercially available TCSPC setups with a pulse burst generation feature, offered by various producers, can be used to implement an RWR excitation scheme. While at present, the use of this feature remains relatively uncommon, the RWR method is a novel protocol that can be applied to the equipment already available in many laboratories worldwide. To facilitate its broad adaptation by the research community, we provide **Supplementary Notes 9** and **10** as a reference manual for the practical realization of RWR TRPL using the Sepia SOM-828 (Picoquant) module.

## Conclusion

To conclude, we propose and demonstrate a Read-Write-Read Time-Resolved Photoluminescence technique which, contrary to the traditional PL decay measurement, takes advantage of excitations with complex pulse sequences to directly probe the material state in multiple time domains ranging from nanoseconds to milliseconds. In contrast to conventional TRPL, our technique probes and shows the photoluminescence from the non-equilibrated state of the material. This novel method makes it possible to visualize directly hidden photoexcited species in a single experiment that are inaccessible by traditional time-resolved PL measurements. This advance is relevant to the field of perovskite photovoltaics and LEDs, in which charge diffusion, recombination and extraction processes are mediated by slow charge trapping and trap depopulation followed by a slower ion migration and even much slower chemical reactions. The developed method allowed us to directly demonstrate trap-filling and de-depopulation processes in $CsPbBr_3$ halide perovskite microplates. From these, we extracted the rate of trap depopulation $k_n$= 1.5×10$^{-10}$ cm$^3$s$^{-1}$ and calculated the concentration of trapped carriers $n_t^0$ = 2.2×10$^{16}$ cm$^{-3}$, which serves as the low limit for the trap density value in this material. We believe that RWR TRPL will become a standard time-resolved technique for studying dynamic luminescent materials, enabling a comprehensive study of performance-limiting processes in emerging photovoltaics, optoelectronic and all-photonic devices.

**Acknowledgements**

The work was supported Swedish Research Council (Grant2020-03530), Crafoord Foundation (Grant 20230552), and NanoLund (Grant 12-2023). J.K. thanks Wenner-GrenFoundation for the postdoctoral scholarship (GrantUPD2022-0132). D.T. and A.P. acknowledge financial support from Russian Science Foundation (project no. 23-72-00031). This project has received funding from the Deutsche Forschungsge-meinschaft (DFG) in the framework of the Special PriorityProgram (SPP 2196) Project PERFECT PVs (Grant424216076).

**Author contributions**

A.M. and I.G.S. conceived the idea and designed the experiments. D.T. and A.P. prepared the samples. A.M. performed measurements with technical support from I.G.S. J.K. modeled the RWR response of the system. A.M, J.K. and I.G.S. contributed to the interpretation of the experimental results. A.M. and I.G.S. wrote the paper with help from Y.V. and with contributions from all authors. The project was supersized by I.G.S. and Y.V.

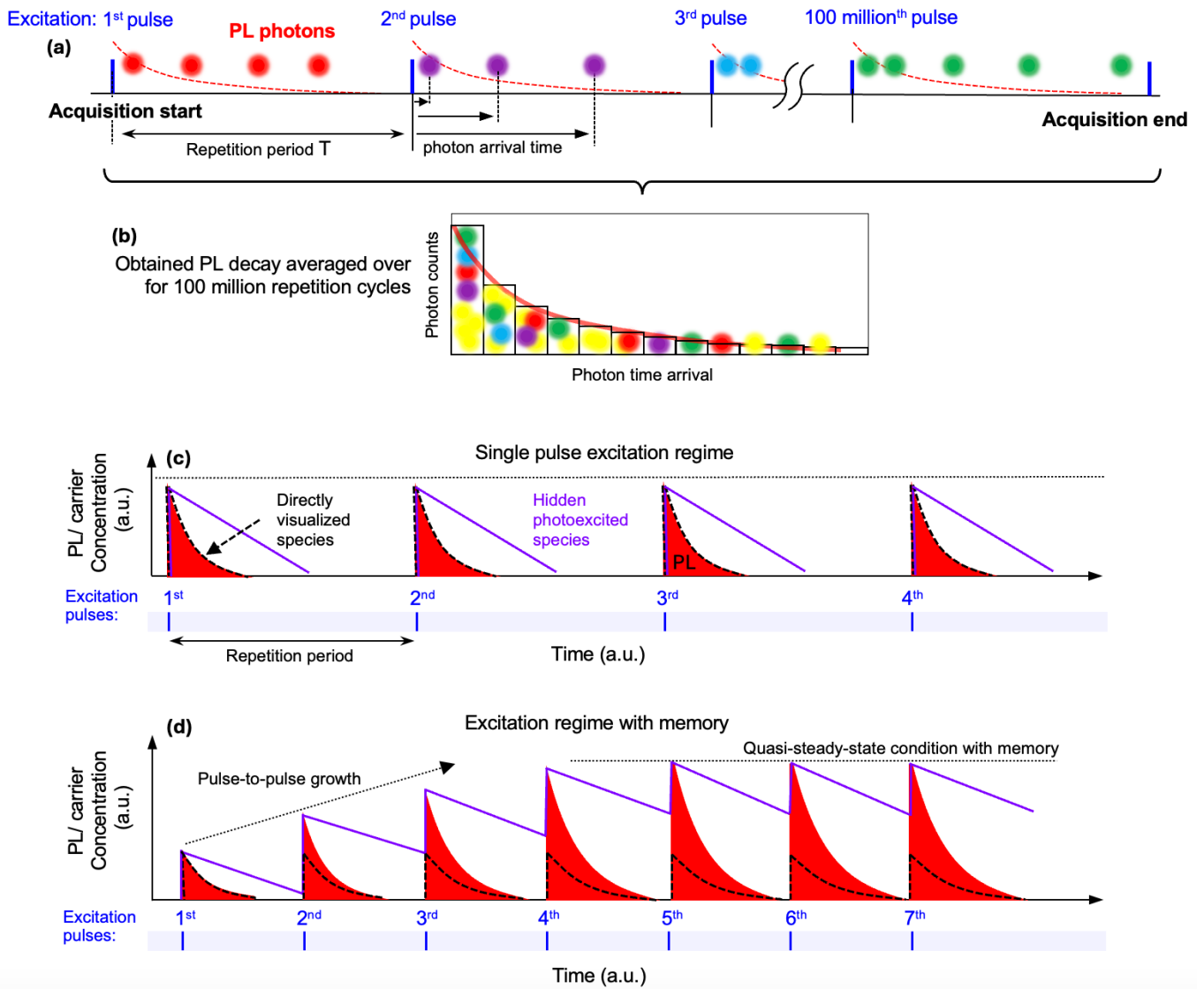

**Figure 1. TCSPC: averaging and consequences for systems with memory. (a)** Time-resolved photoluminescence experiment in the time domain from the start of the acquisition to its end. Each excitation pulse creates PL photons. Photons generated by each pulse are marked with different colors. **(b)** The PL decay histogram is built by sorting the above photons according to their arrival times to the time bins. The information about which pulse creates the particular photons (the color) is lost, which is illustrated by mixing the photons of different colors in each time bin. **(c)-(d)** Imaginary experiment illustrates directly visualized photoexcited species (black dotted line) which follow the PL decay (red) in contrast to hidden photoexcited species (purple line). **(c)** Single pulse excitation regime: the excitation pulse repetition period is much longer than the lifetime of both photoexcited species. The PL response caused by the first pulse is the same as the PL response caused by all further pulses. **(d)** Excitation regime with memory: pulse-to-pulse dynamics leads to the eventual quasi-steady-state condition. PL response is different for the first and last pulses. The final quasi-steady-state condition is denoted here as *quasi steady-state excitation regime with memory*.

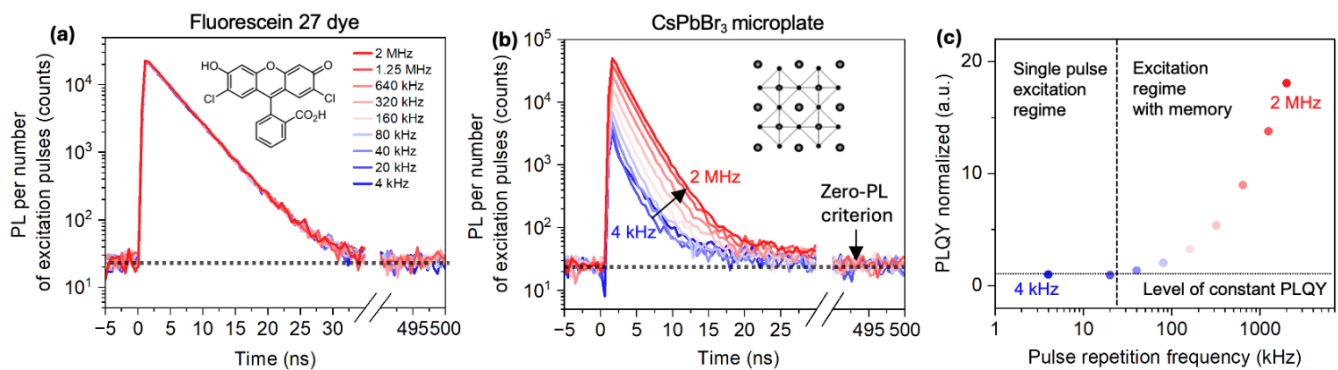

**Figure 2. PL decays measured by different pulse repetition rates.** **(a)** PL decay of the reference dye shows negligible dependence on the repetition rate. **(b)** PL decay of the CsPbBr$_3$ microplate demonstrates a pronounced modulation of the integrated signal (total intensity). **(c)** The integral of the PL decays (shown in b) demonstrates the dependence of PLQY on the pulse repetition rate. The single pulse excitation regime can be seen for repetition rates smaller than around 20 kHz (the region where PLQY does not grow with the repetition rate). The excitation pulse fluence is the same for all graphs and equals 17.6 nJ/cm$^2$.

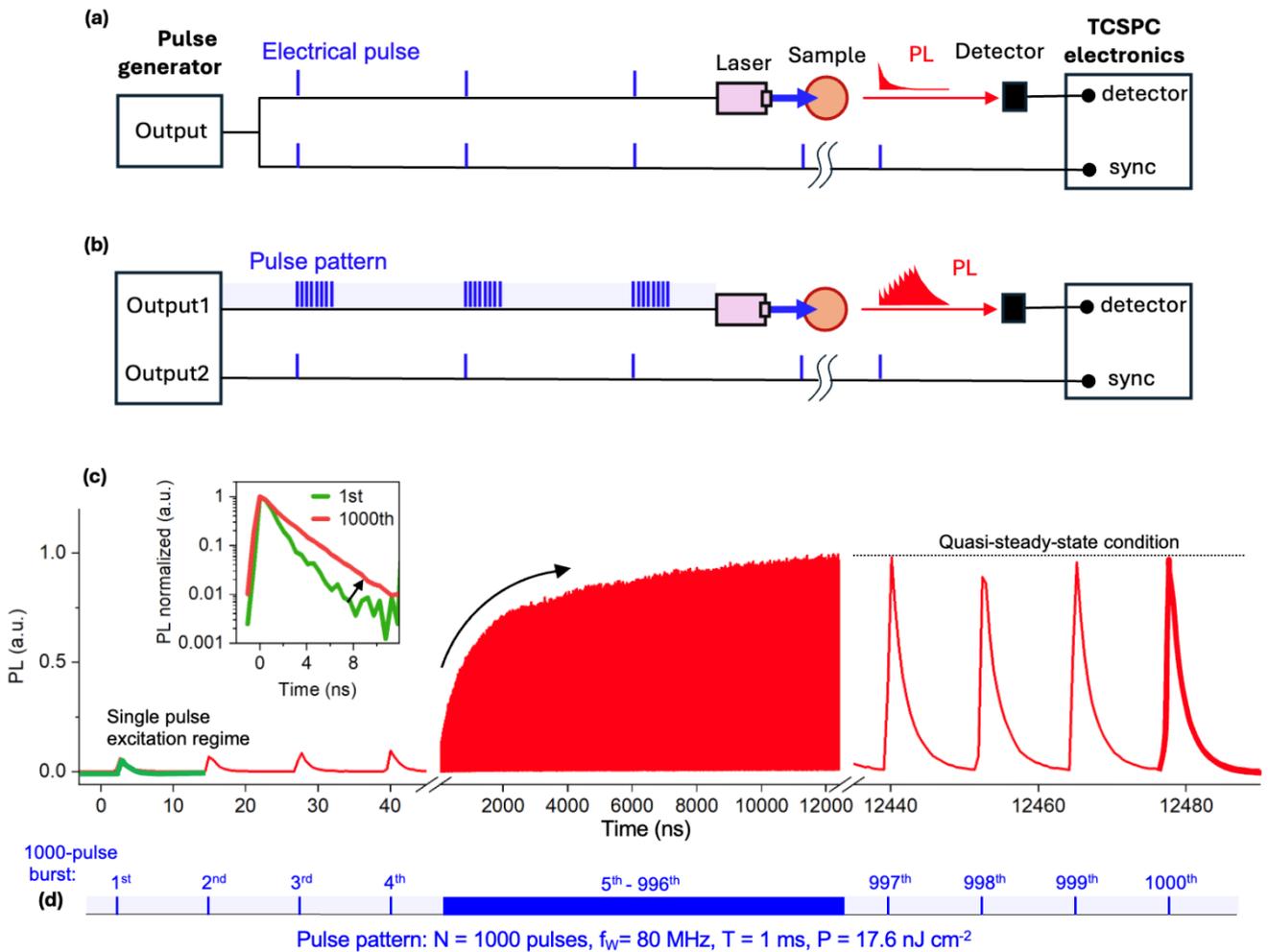

**Figure 3. Pulse burst TCSPC. (a)** Signal processing in the conventional TCSPC with periodic pulsed excitation. **(b)** Single processing in a TCSPC experiment with a periodic pulse pattern (pulse burst) excitation. **(c)** Evolution of the PL intensity and as well the PL decay shape tracked simultaneously for the CsPbBr$_3$ microplate under the condition **(d)** of pulse burst excitation. The responses to the first pulses of the burst (1-4 pulses are shown) are substantially different from those for latest pulses (997 – 1000 pulses are shown). **Inset shows** the normalized PL decays generated by the 1$^{st}$ pulse and the 1000$^{th}$ pulse for comparison.

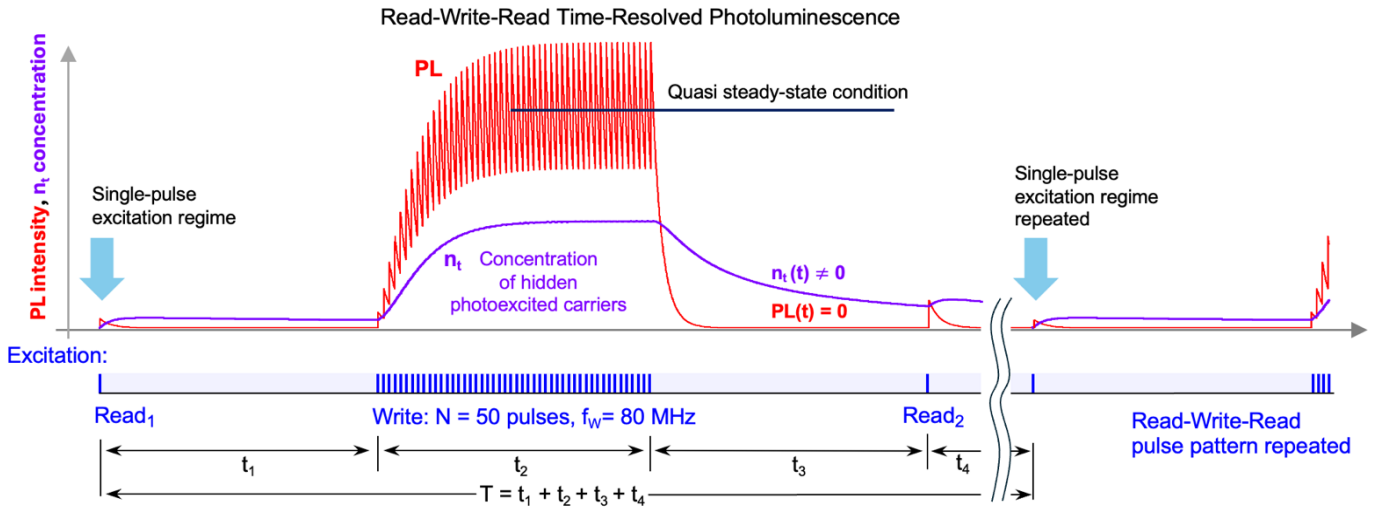

**Figure 4. Read-Write-Read TRPL Technique.** Application of the RWR pulse pattern (bottom) governs a complex photoluminescence response (top). The method is illustrated by the model calculation of the PL response (red) and the concentration of hidden excitations (trapped electrons) in the framework of Shockley-Read-Hall charge recombination model. The first Read$_1$ pulse (blue) excites the sample in the single-pulse excitation regime. Later, the pulse burst (Write) of closely positioned N pulses at a high burst repetition rate $f_W$ (with a total burst duration $t_2$) is applied to substantially change the concentration of hidden photoexcited species (trapped carriers in this model). This may lead to the quasi steady-state condition at the end of the Write burst. Then, the Read$_2$ pulse is applied to probe the state of the sample. The times $t_1, t_2, t_3, t_4$ are the experimental variables defining the experimental design.

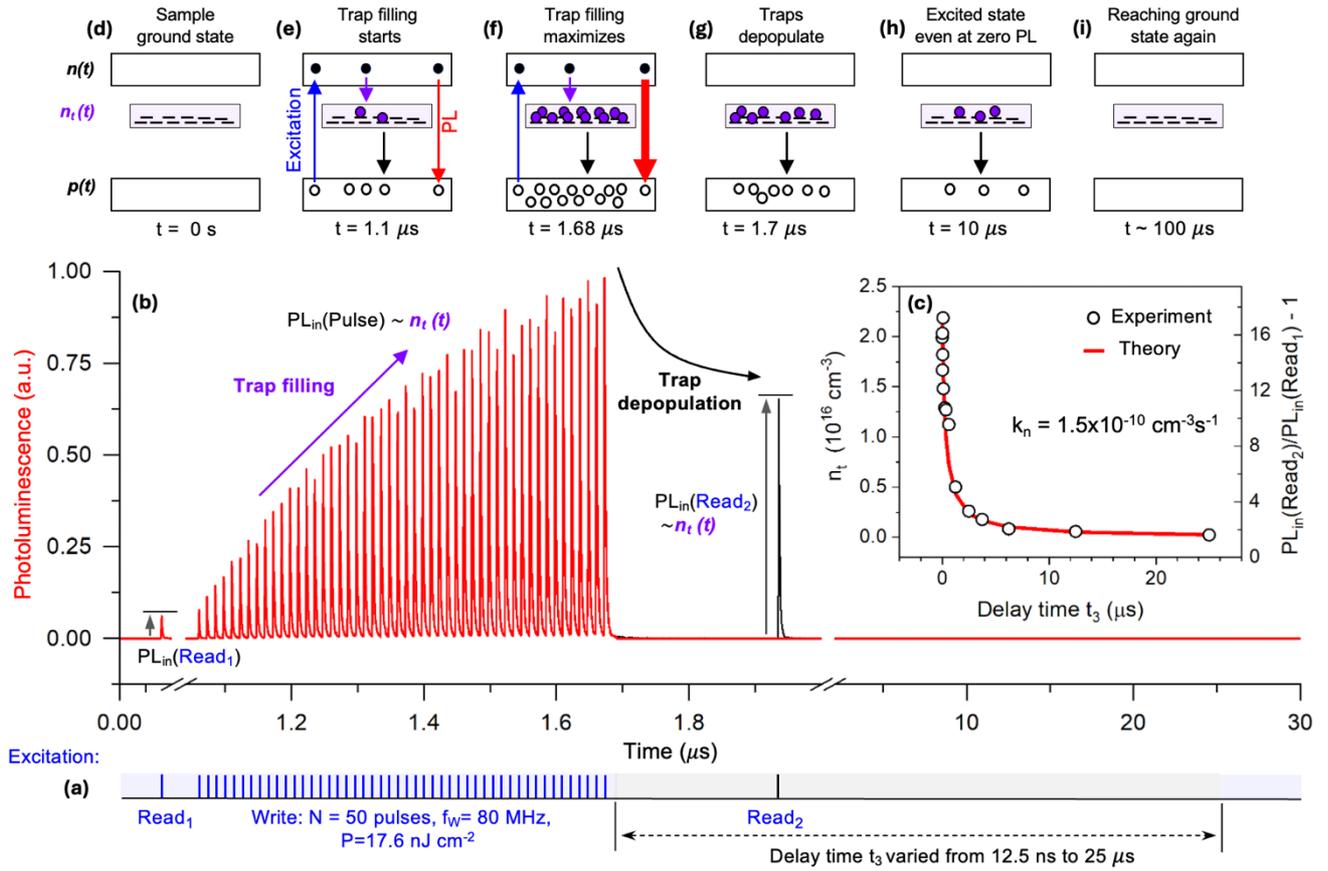

**Figure 5. Visualization of trapped carrier dynamics in CsPbBr$_3$ microplate. (a)** The RWR pulse pattern was applied to the CsPbBr$_3$ microplate. **(b)** PL response to the RWR pulse pattern with time delay $t_3$=0.25 $\mu$s. The information about trap filling and depopulation can be extracted from the initial magnitude of photoluminescence PL$_{in}$(Pulse). **(c)** Ratio between PL$_{in}$(Read$_2$) and PL$_{in}$(Read$_1$) is plotted to extract the filled trap concentration $n_t(t)$ and the rate of trap depopulation k$_n$. **(d)-(i)** Schematic band diagrams describing the RWR experiment. They show the dynamics of the photoexcited carriers: $n(t)$ - concentration of electrons, $p(t)$ – concentration of holes and $n_t(t)$ – concentration of filled traps. Initially the sample is in the ground state **(d)** and all concentrations are zero. Under application of the Write burst, the trap filling process starts **(e)** leading to the pulse-to-pulse PL increase maximizing at t=1.68 $\mu$s. **(f)** After the end of the Write burst, traps start to depopulate (t=1.7 $\mu$s). **(g)** At a time, t=10 $\mu$s, sample is still in the excited state. **(h)** After around t ~ 100 $\mu$s sample reaches the ground state again **(i)**.

# Supplementary information for

# Hidden photoexcitations probed by multi-pulse photoluminescence

Alexandr Marunchenko[1*], Jitendra Kumar[1], Dmitry Tatarinov[2], Anatoly Pushkarev[2], Yana Vaynzof[3,4], Ivan Scheblykin[1*]

*Corresponding author E-mail: shprotista@gmail.com; ivan.scheblykin@chemphys.lu.se;

[1] Chemical Physics and NanoLund, Lund University, P.O. Box 124, 22100 Lund, Sweden

[2] School of Physics and Engineering, ITMO University, 49 Kronverksky, St. Petersburg 197101, Russian Federation

[3] Chair for Emerging Electronic Technologies, Technical University of Dresden, Nöthnitzer Str. 61, 01187 Dresden, Germany

[4] Leibniz-Institute for Solid State and Materials Research Dresden, Helmholtzstraße 20, 01069 Dresden, Germany

## Table of content

# Supplementary Note 1. Setup and RWR TCSPC Technique

## 1.1 Optical Setup

We used the same optical setup as in our previous publications.[1–3] In short, the photoluminescence (PL) was imaged using a home-built wide-field photoluminescence microscope system based on Olympus IX71. The hybrid photomultiplier detector (PMA Hybrid-42 Picoquant) was connected to a time-correlated single photon counting (TCSPC) module (Picoharp 300 Picoquant) for measurements of PL decay kinetics. The PL was excited by a 485 nm diode laser (PicoQuant, with a 200 ps pulse width) through a 40X objective lens. The size of the excitation spot was around 30 μm x 30 μm. The instrumental response function of the TCSPC was approximately 200 ps. For all measurements the pulse fluence was set to 17.6 nJ/cm$^2$ providing the estimated initial carrier concentration of electrons (holes) $n_0$ ($p_0$) = 1.33x10$^{15}$ cm$^{-3}$. The laser was controlled by a Sepia Oscillator Module (SOM 828-D, PicoQuant). This allowed the generating of complex pulse patterns used for PL excitation (see Supplementary Note 8).

## 1.2 TCSPC measurements

**Traditional scheme:**

The traditional TCSPC scheme uses a pulsed laser with a certain repetition rate. Here we performed measurements with different repetition periods from 4 kHz to 2 MHz (Fig.2 of the main text). To conserve the same number of excitation pulses for experiments with different pulse repetition frequencies, we tuned the data acquisition time accordingly. For example, if the acquisition time for 80 kHz pulse repetition rate experiment was 1000 seconds, then the acquisition time for 2 MHz pulse repetition rate experiment was set to 40 s = 1000 s * (80 kHz / 2 MHz).

**Burst and RWR TCSPC:**

We also used bursts of several closely separated laser pulses or other pulse sequences, such as the Read-Write-Read pulse pattern, for the excitation of PL. All of these patterns were created using the SOM 828-D oscillator. The acquisition time for experiments (Fig. 3, Fig. 5) was 10 000 s to achieve a high signal-to-noise ratio.

**Supplementary Note 2. Materials and samples**

**2.1 CsPbBr$_3$ microplates**

**Materials:** Cesium bromide (CsBr, 99.99%, TCI Chemicals), lead (II) bromide (PbBr$_2$, 99.999% trace metals basis, TCI chemicals), dimethyl sulfoxide (DMSO, anhydrous ≥99.8 %, Sigma-Aldrich).

**CsPbBr$_3$ polycrystalline film preparation:** The glass substrates (25 x 25 mm$^2$) were cleaned by sonication in NaHCO$_3$ solution, deionized water, acetone, and 2-propanol for 10 min consecutively, and then exposed to UV-generated ozone for 15 minutes to obtain a hydrophilic surface. Afterwards, substrates were transferred to the dry glovebox filled with N$_2$ gas. The deposition of perovskite films was conducted on the substrates by a single-step spin-coating method at 3000 rpm for 5 minutes. Then, the samples were gradually annealed on a hot plate at temperatures from 50 °C up to 130 °C for 20 min to remove DMSO residues and complete the crystallization.

**CsPbBr$_3$ microplate synthesis:** For the CsPbBr$_3$ microplate synthesis, the temperature difference-triggered growth method was used.[4] For this, a furnace (PZ 28-3TD High-Temperature Titanium Hotplate and Program Regler PR5-3T) was employed to control the temperature during the microplate growth. Briefly, the CsPbBr$_3$ perovskite material was sublimated from a source substrate to the target substrate, separated by a gap of 0.3 cm. The as-synthesized CsPbBr$_3$ polycrystalline film (prepared by the method described above) was used as the source substrate. The target sapphire substrate (20 x 20 mm$^2$) was cleaned by sonication in deionized water, acetone, and 2-propanol for 15 min consecutively. The synthesis was initiated at the furnace temperature of 350 °C. Afterwards, the temperature was increased up to 520 °C for 10 min and kept unchanged for 10 min. As a result, the sublimated CsPbBr$_3$ microcrystals (microplates, microwires) were grown.

**Encapsulation process:** In order to minimize exposure of the obtained perovskite microplates to moisture and oxygen, we used an encapsulation process with a cover glass slip. The process was performed in a nitrogen atmosphere. Initially, a double-sided tape was attached to the surface of a glass cover slip (20x20 mm$^2$ in size and 0.13 mm thick), leaving a free area in the middle of the coverslip. After that, the glass coverslip with the tape was pressed against the sapphire substrate with as-grown perovskite microplates. A two-component epoxy was then used to complete the sealing process. The epoxy was applied at the contact point between the edge of the cover glass slip and the sapphire substrate with the perovskite microplates to hermetically seal the central region. Then, the obtained samples were left in the glove box for 24 hours until the epoxy was completely hardened.

**2.2 Fluorescein 27**

To measure luminescence from the solution of Fluorescein 27 dye, we prepared a diluted solution of Fluorescein 27 in ethanol. The luminescence peak was around 550 nm.

**Supplementary Note 3. Charge carrier recombination model**

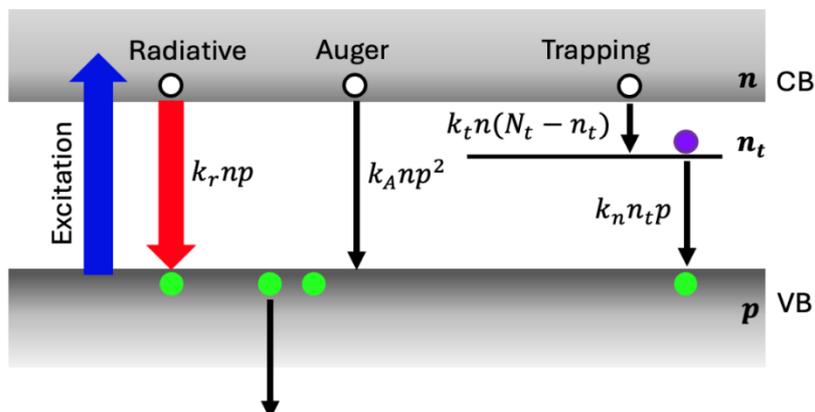

**Figure S1.** Band diagram, transitions and their rates.

We used a Shockley-Read-Hall-based model (Figure S1) to generate the RWR TRPL response in Figure S2 and Figure 4 of the main text. The model includes the first order non-radiative recombination via traps, the second order radiative recombination, and the third order Auger recombination. It is described by a system of equations:[1,5]

$$\frac{d}{dt}n(t) = G(t) - k_t(N_t - n_t)n - k_r np - k_a np^2 \qquad \text{(S1)}$$

$$\frac{d}{dt}n_t(t) = k_t(N_t - n_t)n - k_n n_t p \qquad \text{(S2)}$$

along with the condition of charge neutrality

$$n(t) + n_t(t) = p(t) \qquad \text{(S3)}$$

Here, n and p, are the electron and hole densities in conduction and valence band respectively. $n_t$ is the trapped electron density. G(t) is the charge carrier generation rate, and $N_t$, $k_t$, $k_n$, $k_r$, and $k_a$, are the trap density, electron trapping rate, non-radiative recombination rate constant, radiative recombination rate constant and Auger recombination coefficient respectively.

Note, that excitons are not included to this model because of their low fraction compared to free charge carriers. To estimate the fraction of exciton we used the Saha-Langmuir equation:

$$\frac{x^2}{1-x} = \frac{1}{n_{ex}}\left(\frac{2\pi\mu kT}{h^2}\right)^{3/2} e^{-\frac{E_b}{kT}}$$

Where $x$ is the fraction of free charge carriers, n is the density of excitation (we consider $n_{ex}$ = 1.33x10$^{15}$ cm$^{-3}$), $\mu$ is the reduced mass of exciton (approximated to be 0.25 $m_e$), h is the Planck's constant, the exciton binding energy is chosen to be $E_b$ = 40 meV for CsPbBr$_3$ metal halide perovskites studied at room temperature ($kT$ = 25 meV).

After solving the Saha-Langmuir equation we obtain that the fraction of excitons (1-x) equals to 0.5%. This means that PL intensity is defined by the recombination of free electrons and holes rather than excitons and given by:

$$PL(t) = k_r n(t) p(t) \tag{S4}$$

For more details see also our previous publications.[1,5]

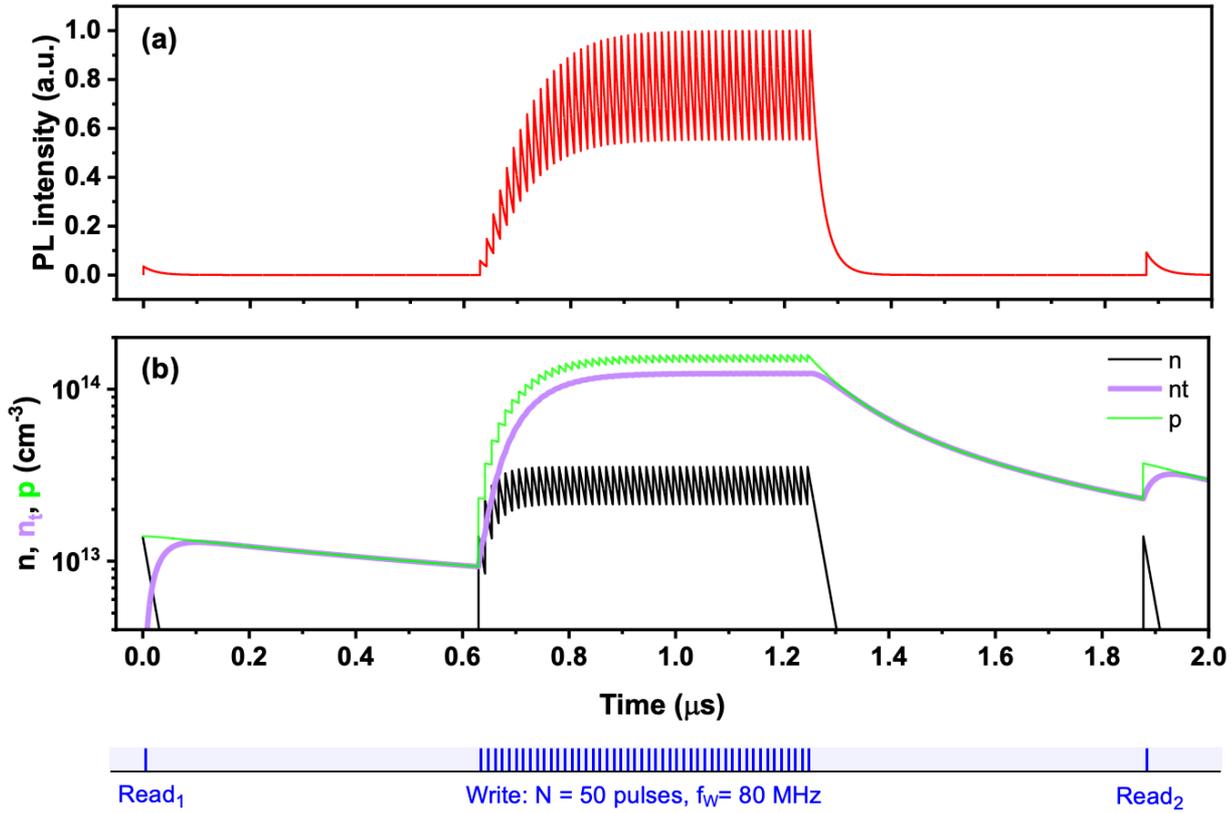

**Figure S2**. **(a)** The same data as shown in Fig. 4 of the main text. The RWR excitation pulse pattern is shown at the bottom. **(b)** The modeled evolution of the electron density in conduction band (n), the hole density in valence band (p), and the trapped electron density ($n_t$).

**Table S1.** Model parameters used to calculate the PL response in Figure S2

| | |
|---|---|
| $k_t N_t$ | 4.0x10$^7$ s$^{-1}$ |
| $N_t$ | 1.75x10$^{16}$ cm$^{-3}$ |
| $k_n$ | 5.93x10$^{-8}$ cm$^3$s$^{-1}$ |
| $k_r$ | 1.73x10$^{-11}$ cm$^3$s$^{-1}$ |
| $k_a$ | 2.84x10$^{-29}$ cm$^3$s$^{-1}$ |

**Supplementary Note 4. Details of the RWR TRPL experiment on CsPbBr$_3$**

The figure below shows the same data as Fig.5 in the main text but in a different presentation. Here that all PL responses are plotted together. The difference is observed in the PL response to the Read$_2$ pulse as a function of t$_3$. The responses to the Read$_1$ pulse and the Write burst are not changing. Panel (e) shows that the PL decay excited by Read$_2$ pulse is longer for small t$_3$ in comparison with the large t$_3$. The PL response for large t$_3$ is very similar to the PL response to the Read$_1$ pulse.

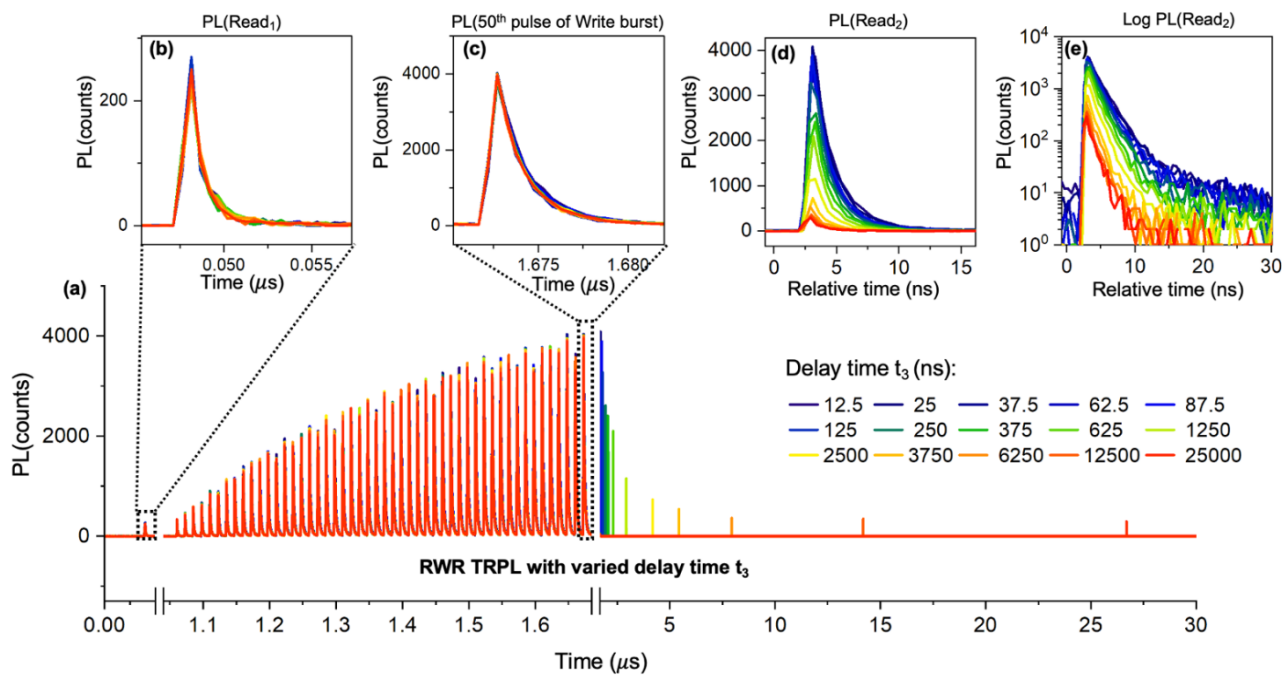

**Figure S3.** Extra details of the experiment shown in the main text in Fig. 5. **(a)** The RWR pattern (see also Fig.7) with t$_1$ = 1 μs, N = 50, f$_W$ = 80 MHz, T = 1 ms and a pulse fluence equal to 17.6 nJ/cm$^2$ is applied to the CsPbBr$_3$ microplate. The delay time t$_3$ is varied from 12.5 ns to 25 μs (see the legend). PL responses for all different delay times t$_3$ are plotted together on the same graph. Almost no difference is visible in the PL responses to Read$_1$-Write pulse patterns **(b,c)**, while the PL response to the Read$_2$ pulse is dependent on t$_3$ **(d,e)**.

## Supplementary Note 5. Time dependence of the memory effect in CsPbBr$_3$

The dependence of the ratio of the time-integrated responses to Read$_2$ and Read$_1$ pulses presented in the figure below is fitted with the three-exponential function $1 + A_1 e^{-t/\tau_1} + A_2 e^{-t/\tau_2} + A_3 e^{-t/\tau_3}$, where $A_1$=21, $A_2$=5.86, $A_3$=0.93 and the characteristic decay times $\tau_1$=0.384, $\tau_1$=2.08 and $\tau_1$=25.2 $\mu s$ respectively. The constant 1 is needed because at the end of the relaxation, the PL response to the Read$_2$ pulse should become the same as the PL response to the Read$_1$ pulse.

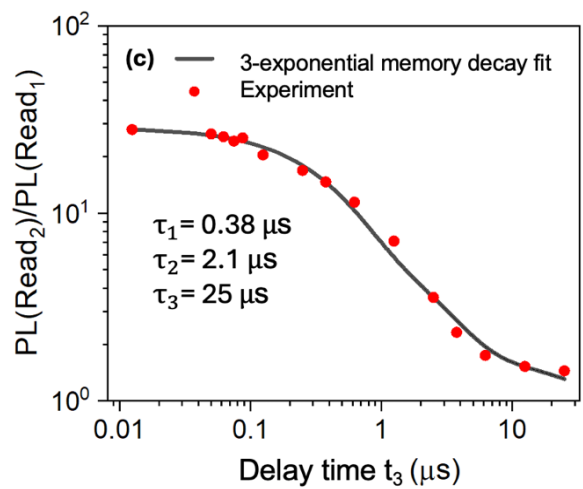

**Figure S4.** The dependence of the ratio of the time-integrated PL(Read$_2$) to PL(Read$_1$) as a function of the $t_3$ delay time. The solid line is the fit by a 3-exponential function.

## Supplementary Note 6. PL decay in the case of strong photodoping

PL intensity is given by Eq. S4. In the case of strong photodoping, the decay of n (minority charge carrier in our model) is much faster than the decay of p (majority charge carrier). Thus, during an experimentally observed PL decay, p does not change much, while the decay of n causes the observed PL intensity to decay. So, at first approximation, PL(t) decays as the minority charge carrier population. For our study, it has an important consequence: when we see that PL intensity has decayed e.g. 10 times between two excitation pulses in a Write pulse burst, it means that the concentration of electrons at the moment of arrival of the next excitation pulse is also approximately 10 times smaller than it was created by the previous excitation pulse.

## Supplementary Note 7. Analytical approximation for the decay of trapped carriers

At a low excitation condition, PL quantum yield is low, and thus, we can disregard the radiative and Auger recombination terms in Eq.S1. The generation rate G is zero because we do not excite the semiconductor and want to calculate how the concentration of already existing trapped electrons evolves over time.

Then we re-write Eq.S1 and S2:

$$\frac{dn}{dt} = -k_t(N_t - n_t)n \tag{S5}$$

$$\frac{dn_t}{dt} = k_t(N_t - n_t)n - k_n n_t p \tag{S6}$$

or

$$\frac{dn_t}{dt} + \frac{dn}{dt} = -k_n n_t p \quad (S7)$$

Since we also assume the case of strong photodoping (n << p), then $n_t \approx p$ due to charge neutrality (Eq. S3), thus Eq S7 becomes

$$\frac{dn_t}{dt} + \frac{dn}{dt} = -k_n p^2 \quad (S8)$$

Using again the charge neutrality condition (Eq. S3) for the left part of the Eq. S8 we obtain:

$$\frac{dp}{dt} = -k_n p^2 \quad \text{or} \quad -\frac{dp}{p^2} = k_n \, dt \quad (S9)$$

This equation can be analytically integrated:

$$\int_0^t \frac{dp}{p} = -\int_0^t k_n \, dt$$

$$\frac{1}{p(t)} - \frac{1}{p(0)} = k_n t$$

$$p(t) = \frac{p(0)}{1 + k_n p(0) t} \quad (S10)$$

Where t=0 is the time moment we start monitoring the decay of p(t).

Because $n_t \approx p$, the same time dependence is valid for $n_t$ too:

$$n_t(t) = \frac{n_t(0)}{1 + k_n n_t(0) t} \quad (S11)$$

**Supplementary Note 8. Comparison to other methods**

It is noteworthy that apart from the RWR method proposed here, fully optical methods capable of directly probing the population and relaxation of trapped charge carriers in semiconductors are scarce. The pump-probe transient absorption (TA) spectroscopy is the most well-known and universal method. The main advantage of TA is that, in theory, the concentration of any photoexcited species absorbing light can be probed over a very broad time range and with an excellent time resolution limited only by the laser pulse width.[6–8] However, the interpretation of TA results is often complicated by the absence of distinct spectrally separated features of each of the different excited species due to the broadness and significant overlap in their spectra.[9,10] For example, the signal from trapped charge carriers can be masked by a much stronger signal from free charge carriers. TA is also rather expensive and technically demanding, requiring significant know-how and expertise.

Other contactless methods are the time-resolved microwave conductivity and THz time-resolved spectroscopy. While theoretical modeling of the measured signals can be used to extract certain parameters related to charge trapping and trap depopulation, in most cases, their direct study is not possible.[11–14]

The introduction of electrical contacts to semiconducting samples makes it possible to use additional methods where electrical current or voltage is measured in response to light irradiation, and information regarding charge trapping and trap depopulation can be obtained. However, the presence of contacts in many cases is a drawback since they may induce processes at the interfaces and induce capacitive effects, which distort the signatures of trapped charge carriers, introducing both limitations on and even errors in data interpretation.[15–17] A combination of optical pump-probe measurements with electrical measurements can provide a direct measure of trap state population with a great temporal resolution, keeping, however, the drawback of having contacts.[18]

**Supplementary Note 9. Guide for realizing RWR TRPL scheme with Sepia SOM 828-D**

**1. Realization of TRPL with pulse burst excitation and the Read-Write-Read scheme using TCSPC electronics.**

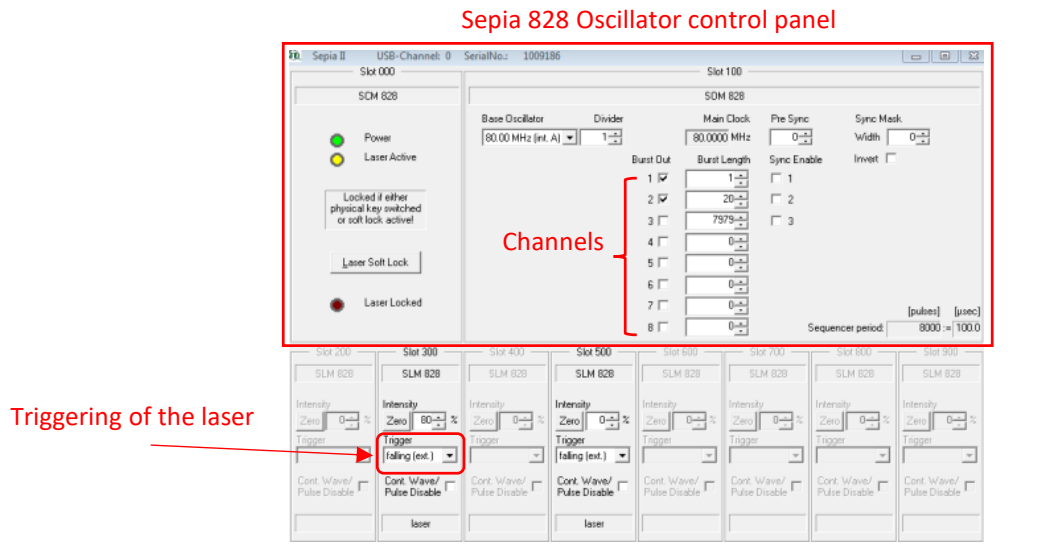

| Ch | Pulses | Note | Output (yes/no) |
|----|--------|------|-----------------|
| 1 | 1 | Connect this output to sync 0 of the TCSPC system | **YES** |
| 2 | 20 | Burst | **YES** |
| 3 | 7979 | Gap (not connected) | NO |
| 4 | 0 | | NO |
| 5 | 0 | | NO |
| 6 | 0 | | NO |
| 7 | 0 | | NO |
| 8 | 0 | | NO |

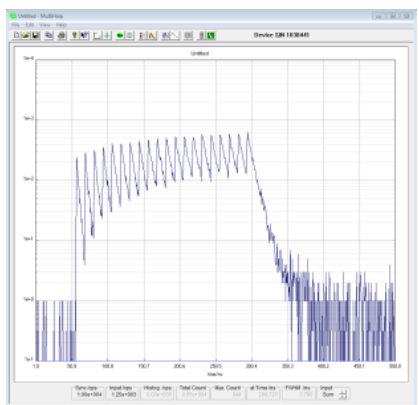 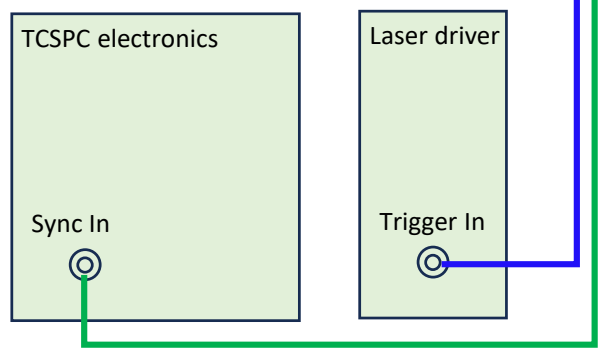

**Figure S5**. Realization of a TRPL measurement with excitation by pulse bursts using SOM 828 (PicoQuant).

In this section, we will give a practical description of how to adapt a standard TCSPC setup for measurements with the excitation by pulse bursts. We will use as an example a PicoQuant system installed in our laboratory consisted of a PicoHarp150 (or PicoHarp300) counting device, picosecond diode laser driven by Sepia 828 with Sepia oscillator module (SOM) SOM 828 or SOM 828-D.

The pulsed laser must be externally triggered by the oscillator, which is able to produce pulse bursts of electrical pulses. In the software, one chooses "falling external" as the triggering regime of the laser. The oscillator output Channel 1 (Fig.S5) is connected to the laser trigger input. In this configuration, the laser operates according to the oscillator's settings.

SOM 828 has 8 output channels. In the software, one can specify how many pulses each channel should produce. These channels provide signals in a sequence ch1, ch2, ch3..., and ch8, which is constantly repeated with the period determined by the total number of pulses in the sequence and the base frequency. The software can specify the base frequency by setting a divider to 80 MHz. The base frequency is the repetition rate in the pulse burst ($f_w$), setting the time distance between the pulses. This distance at the base frequency of 80 MHz equals 12.5 ns. More details can be found on the producer's website:

[https://www.picoquant.com/products/category/picosecond-pulsed-driver/pdl-828-sepia-ii-computer-controlled-multichannel-picosecond-diode-laser-driver#description].

Fig. S5 demonstrates the realization of a TRPL experiment with excitation by periodic pulse bursts (number of pulses in the burst N = 40, internal pulse repetition rate $f_w$ = 80 MHz ) and 10 kHz total repetition rate (equivalent to the period T=100 µs) using SOM 828. Notably, one cannot directly realize the Write-Read TRPL and Read-Write-Read TRPL schemes discussed in the article using SOM 828. This is because SOM 828 does not allow connection of the laser trigger input with more than one output channel of the oscillator, which is required for the Write-Read (2 channels) and Read-Write-Read (3 channels) schemes.

However, we still succeeded in combining the signals from two output channels of the SOM 828 oscillator module. For that, we connected two coaxial cables with T-shaped BNC connectors. With this, we were able to realize the Write-Read TRPL scheme used in the recently published article.[3] However, such direct connection of the cables leads to signal attenuation and reflections on the junctions. Therefore, there is no guarantee that this simple solution will work for another setup. In our case, combining more than 2 channels directly was impossible. However, SOM 828, with 2 channels electrically combined, is already a good starting option.

In contrast to SOM 828, the newer version of the oscillator SOM 828-D allows combining signals from different channels into one output using the software. This output is then used to trigger the laser. Hence, one can perform RWR TRPL measurements. Fig. S6 and Fig. S7 show an example of the realization of the RWR TRPL experiment with 20-pulse burst excitation, internal burst frequency 80 MHz, and 1 kHz total repetition rate (equivalent to the period T=1 ms).

Sepia 828D Oscillator control panel

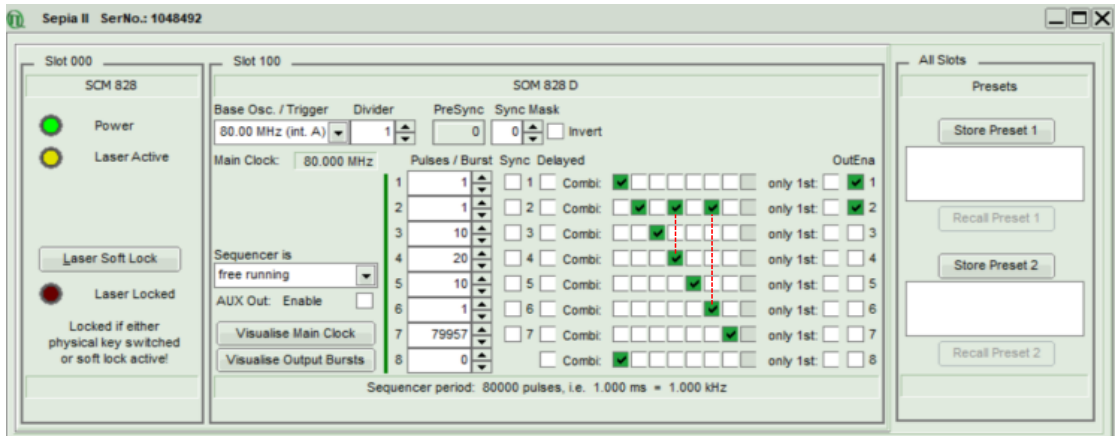

| Ch | Pulses | Note | Output (yes/no) |
|---|---|---|---|
| 1 | 1 | Connect this output to sync 0 of the TCSPC system | **YES** |
| 2 | 1 | Read1 pulse | **YES** |
| 3 | 10 | Gap (not connected) | NO |
| 4 | 20 | Write Burst | NO |
| 5 | 10 | Gap,(not connected) | NO |
| 6 | 1 | Read2 pulse | NO |
| 7 | 79957 | Gap (not connected) | NO |
| 8 | 0 |  | NO |

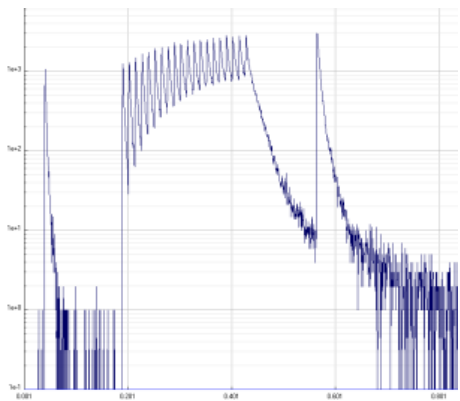
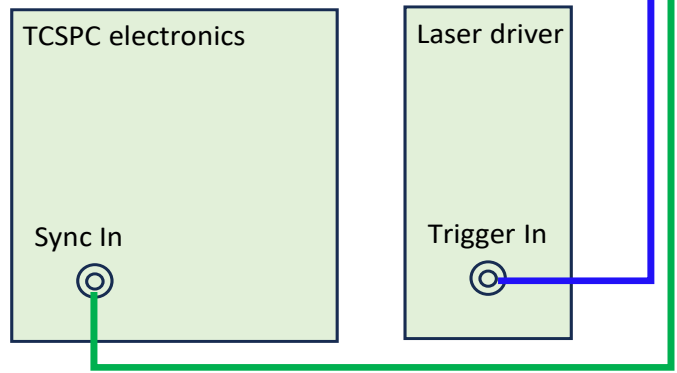

**Figure S6**. Realization of a Read-Write-Read scheme using SOM 828-D (PicoQuant).

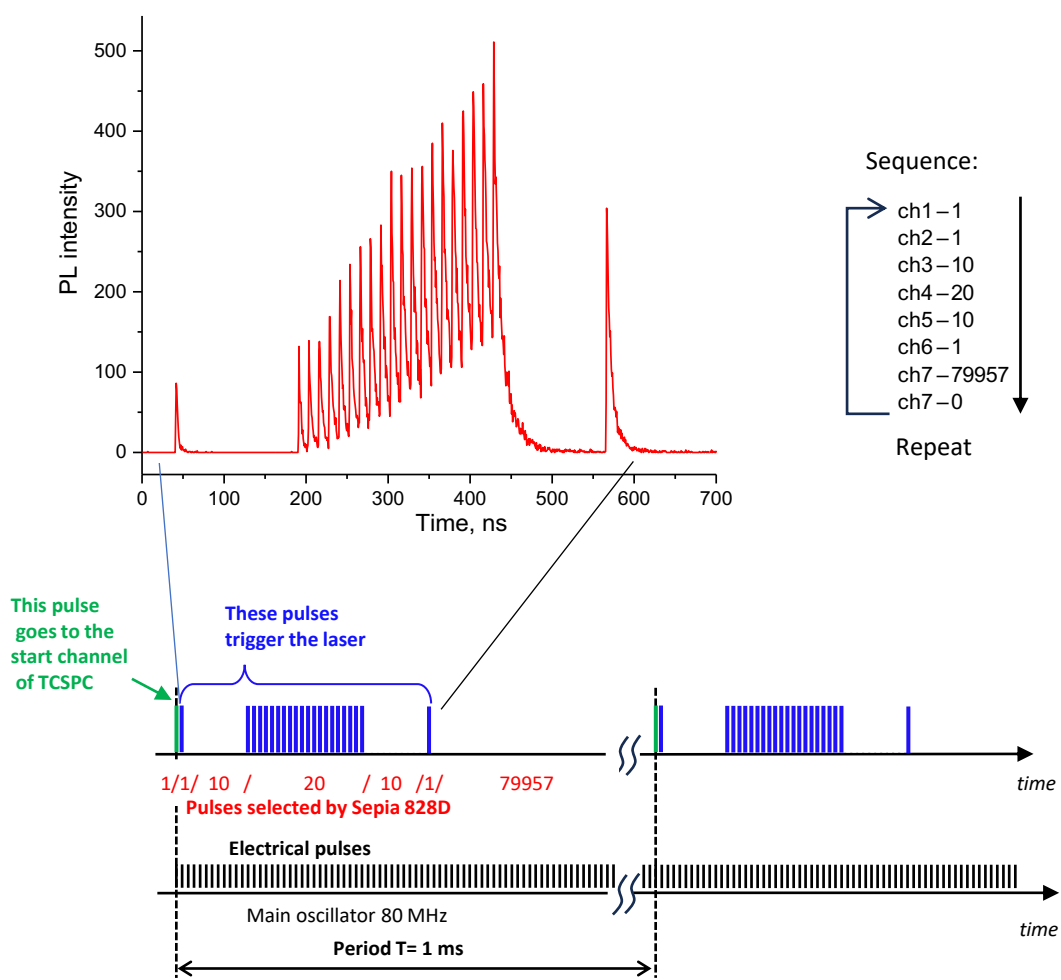

**Figure S7**. Additional description of the Read-Write-Read scheme introduced in Fig.S6.

**Supplementary Note 10. Possible photon count rates in TCSPC measurements.**

It is important to mention practical issues concerning the possible photon count rates in TCSPC measurements when the repetition rate of excitation (single pulse or any pulse pattern like a pulse burst) is very low (10 kHz range and lower).

It is common to consider that the so-called "5% rule" (less than 5% probability to detect photon per one excitation pulse) should always be applied during TCSPC data acquisition to avoid the pile-up effect (when the detector misses photons that arrive too closely to one another in time (Fig.S8a)) disturbing the decay curve.[19,20] However, if one would like to follow this rule for a 1kHz excitation repetition rate (the equivalent of 1 ms period) like in the RWR TR PL experiment shown in Fig.S6 and Fig.S7), the possible photon count rate should be limited to 1000 * 5% = 20 cps only. With this count rate, accumulating data with a decent signal-to-noise ratio will take a very long time. However, we obtained a very good signal-to-noise ratio with an acquisition time of 100 s only in the decay shown in Fig.S7, which was acquired with a photon count rate of 700 cps, which is 70% of the start rate. Remarkably, the measured TRPL

**(a) High repetition rate**

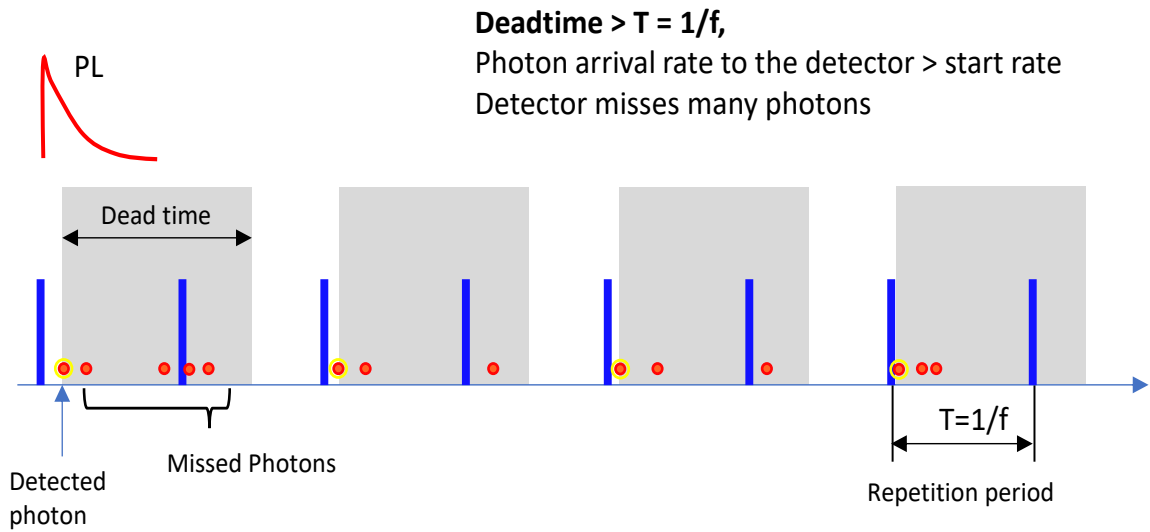

**Deadtime > T = 1/f,**
Photon arrival rate to the detector > start rate
Detector misses many photons

For correct measurements **Photon Count Rate << Start Rate**
Usual rule: **Photon Count Rate < 5% Start Rate**

**(b) Low repetition rate**

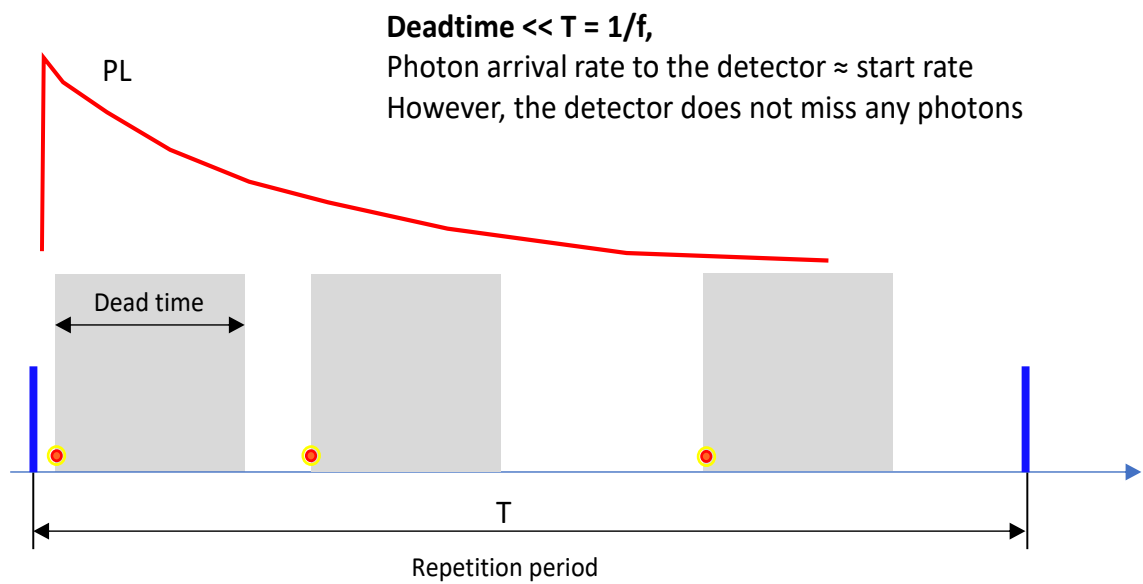

**Deadtime << T = 1/f,**
Photon arrival rate to the detector ≈ start rate
However, the detector does not miss any photons

**Measurement can be correct even when Photon Count Rate is comparable with the Start Rate. The 5% rule can be ignored (with care).** One needs to consider the number of photons within the deadtime at any time moment of the PL decay

**Figure S8**. Photon counting rate in TCSPC experiments.

response does not contain any artefacts that one may expect to appear due to the high detection count rate.

To understand why no issues arise during such an experiment, we need to recall the operation principles of the TCSPC setup. Contrary to the very first generations of TCSPC electronics, which could detect only one stop event per start event (the true start-stop scheme), the modern TCSPC setups are able to count many photons and determine their arrival times relative to the start pulse. The only limitation here is the detection dead time.

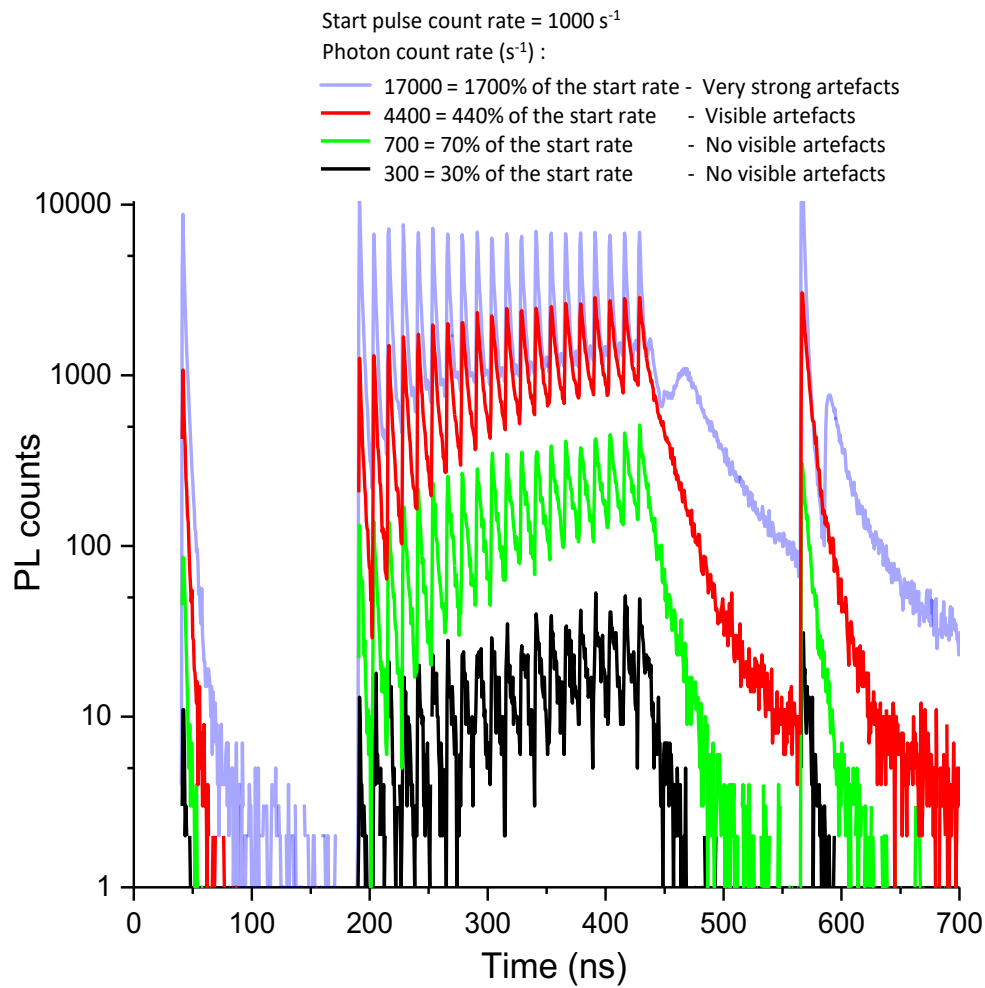

**Figure S9.** Dependence of the measured PL response in the RWR scheme on the photon count rate.

If the dead time is longer than the repetition period (T) of the excitation pattern, then not more than one photon per start event can be detected, and the 5% rule applies. The dead time of the detector is usually around 100 ns. So, for this dead time and the repetition rate of excitation larger than 10 MHz (equivalent to 100 ns period), the "5% rule" is applied (Fig. S8a). However, several photons can be safely detected when the dead time is much smaller than the period T (repetition frequency in the kHz range and lower) (Fig.S8b).

> *Formally, when dead time << T, the conditions for the entire PL decay curve to be correct can be formulated as follows: the probability for a photon to arrive within the time window equal to the dead time should be small (~ 5%) for any position of the time window within the full repetition period T.*

**In practice, this leads to an important consequence: at a low excitation repetition rate, photon count rates substantially higher than 5% of the repetition rate can be used without disrupting the PL response.**

It is impossible to give a simple recipe for the count rate limit because the limit depends on the shape of the PL response. Our recommendation is simply to start measuring at a low photon count rate, then increase it, and follow the PL response until visible distortions of the PL shape are observed. Then, decrease the rate, for example, by a factor of 10 to be safe. Therefore, one can find an optimal regime of measurement at a low repetition rate of the excitation when counting substantially more photons than allowed by the "5% rule" still without any PL response distortion.

In the example below, we will show how choosing a high but still safe count rate allows the use of 1-2 orders of magnitude faster signal collection than if one follows the "5% rule". Fig. S9 shows the RWR TRPL response measured for 1kHz excitation repetition rate (the equivalent of T=1 ms period) for different OD filters installed before the detector. Changing OD allowed for a change in the number of photons arriving at the detector. The number of PL photons reaching the detector is approximately 10 times different from one curve to the next.

Note that the lowest count rate of 300 cps (Fig.S9 black curve) is already 30% of the start rate (1000 Hz), and this count rate cannot be reduced because most of these counts are just dark counts of the detector. Namely, 300 cps consists of approximately 260 cps of noise (mostly electrical noise) and 40 counts of the PL photons.

However, these 300 cps (or 0.3 photons per period T=1 ms) do not lead to any pile-up effect despite violating the "5% rule". This is because they are equally spread over time, and the average distance between them (1 / 0.3 ms-1 = 3.3 ms) is much larger than the 100 ns detection deadtime of our TCSPC system. So, the background photons can be simply ignored.

Upon removing one OD=1 filter, we increase the amount of PL photons by a factor of 10 (Fig.S9 green curve). The count rate reaches 700 cps (70% of the start rate), among which PL photons give 440 cps or 0.44 photons per repetition period. Due to the shape of the PL response (Fig. S9), most of these photons are concentrated within an approximately 300 ns window. So, we have 0.44 photons per 300 ns or 0.15 per the dead time (100 ns). This is a rather safe count rate, which is why we do not have any visible artefacts in the obtained PL response.

When the PL intensity is increased 10 times more (Fig. S9 red curve), the number of photons per dead time reaches 1.5, which results in visible artefacts, making the PL response (red curve) differ in shape from the green curve. A further 10 times increase makes the artefacts very strong (Fig.S9 light violet curve). These examples demonstrate how the artefacts may look and how to choose a high but still safe count rate in TCSPC experiments when the repetition period is much larger than the detector dead time.